\documentstyle[aps,multicol,prl]{revtex}
\begin{document}
\newcommand{\be}{\begin{equation}}
\newcommand{\ee}{\end{equation}}
\newcommand{\ba}{\begin{eqnarray}}
\newcommand{\ea}{\end{eqnarray}}
\title{Anomalous Exponent of the Spin Correlation Function of  
a Quantum Hall Edges} 
\author{H. C. Lee}
\address{
Asia Pacific Center for Theoretical Physics, Seoul, Korea  }
\author{S.-R. Eric Yang}
\address{
Department of Physics, Korea University, Seoul 136-701, Korea  }
\draft
%\date{\today}
%%%%%%%%%%%%%%%%%%%%%%%%%%%%%
\maketitle
\begin{abstract}
\indent  The charge and spin 
correlation functions of partially spin-polarized edge
electrons of a quantum Hall bar are studied using 
effective Hamiltonian and bosonization techniques. 
In the presence of the Coulomb interaction between the edges with opposite
chirality we find a different 
crossover behavior in spin and charge correlation functions.  
The crossover of the spin correlation function 
in the Coulomb dominated regime is 
characterized by an anomalous exponent, which originates from
the finite value
of the effective interaction for the spin degree of freedom in the long 
wavelength limit.  The anomalous exponent may be determined by
measuring nuclear spin relaxation rates in a narrow quantum Hall bar
or in a quantum wire in strong
magnetic fields.
\end{abstract}
%
%\tableofcontents
%\vskip 1.0cm
\pacs{{\rm PACS numbers}: \hspace{.05in} 73.20.Dx, 73.20.Mf,  73.40.Hm }
%\pagebreak
%%%%%%%%%%%%%%%%%%%%%%%%%%%%%
\begin{multicols}{2}
Edges of
quantum Hall (QH) bars \cite{W1,W2,W3,MYKGF} have attracted a
considerable attention recently.  This is because they provide clean
experimental verifications of many interesting theoretical predictions of 
Luttinger liquids \cite{MUW,CPW}.
Recent investigations have shown that
the long-range Coulomb interaction brings new effects 
into these systems \cite{FB,OF,moon96,LY}.
In {\it spin-polarized} systems it introduces a 
cross-over from a Luttinger/Fermi liquid (power law) regime 
to a regime where the inter-edge Coulomb interaction dominates.
A good example is the tunneling conductance at filling factor $\nu \le 1$ 
between the quantum Hall edges of opposite chirality.  
It is described by \cite{moon96}
\be
G \sim \cases{ (\frac{T_0}{T} )^{2(1-1/\nu)}, T > T_0 \cr
       (\frac{T_0}{T} )^{2} \exp \{-\frac{2 \sqrt{2 \chi}}{3 \nu}
     ( \ln \frac{T_0}{T})^{3/2}  \}, T < T_0, \cr}
\ee
where $T_0$ is the cross-over temperature scale. 
Similar cross-over exists in the CDW correlations \cite{FB,schulz93,zulicke96}
\begin{eqnarray}
\label{cdw}
C(x)\sim \cases{   (x/\ell)^{-2/\nu}, \;\;x<W \cr
\cos(2k_Fx) \exp(-\frac{2}{\nu}\sqrt{\alpha\ln(x^2 \beta)}), \;x>W, \cr} 
\end{eqnarray}
where $\ell$, $W$, and $k_F$ are the magnetic length, the width of the Hall
bar, and the Fermi wavevector.

In {\it partially} spin-polarized
QH edges correlation functions are expected to exhibit interesting
properties.  Since the guiding center of
the single particle wavefunction depends on the value
of the wavevector, the  
wavefunctions of spin-up and -down electrons at the Fermi wavevectors
are spatially separated. 
Moreover, spin and charge correlation functions are expected to
behave differently since  
spin and charge separate 
\cite{LY}.
We have investigated how spin and charge correlation
functions behave as the transverse width of the Hall bar
changes.  We find 
that, although charge correlation functions are similar to
the result given in Eq.(\ref{cdw}),
the spin correlation functions
acquire {\it anomalous}
exponents $ \alpha_{2 k_F}$ and $\alpha_0$.
The imaginary part of transverse  spin-spin  correlation functions $ S(\Omega) $
consist of inter- and intra-branch 
terms:
\begin{eqnarray}
\label{answer}
S_{{\rm inter}}(\Omega) \propto \cases{ 
 \,\Omega,\;\;
  \Omega > \Omega_{{\rm cr}} \cr
\Omega^{\alpha_{2 k_F}}\,\,
e^{-\beta_{2 k_F} [\ln \frac{v_\rho}{|\Omega|d} ]^{1/2} },
 \;\; \Omega < \Omega_{{\rm cr}}, \cr}
 \end{eqnarray}
\begin{eqnarray}
%\label{answer}
S_{{\rm intra}}(\Omega) \propto \cases{ 
 \,\Omega,\;\;
  \Omega > \Omega_{{\rm cr}} \cr
\Omega^{1+\alpha_0}\,\,
e^{-\beta_0 [\ln \frac{v_\sigma}{|\Omega|d} ]^{1/2} },
 \;\; \Omega < \Omega_{{\rm cr}}. \cr}
 \end{eqnarray}
Here $v_{\sigma (\rho)}$, $d$, $\beta_{0 (2 k_F)}$, 
and $\Omega_{{\rm cr}}$ are 
some appropriate velocity, 
length, dimensionless constant, and crossover frequency. 
In the low energy limit the inter-branch
contribution dominates over the intra-branch term since
$\alpha_{2 k_F} < 1+ \alpha_0$.  However,
the prefactor of the inter-branch term 
turns out to decay very rapidly as a function of the width of the Hall bar.
Consequently, the intra-branch term is more relevant in
wide Hall bars.  The anomalous 
exponent $\alpha_0$ goes to zero in the limit where the
width of the Hall bar goes to infinity, and an expression similar
to Eqs. (1) and (2) is recovered.
For a narrow Hall bar or quantum wire in strong magnetic fields
the anomalous exponent $\alpha_{2 k_F}$ is significant
since the edge separation
is not negligible compared to the transverse width.   
The presence of the anomalous exponent in the transverse spin 
correlation function
may be verified experimentally by measuring
nuclear spin relaxation rates in a quantum wire in strong magnetic fields.
In both expressions for the spin correlation function the 
power laws in $\Omega$  
originate from the spin sector.  
The anomalous exponent emerges because the difference between spin-up and down
electron
wavefunctions at the respective Fermi wavevectors   
makes 
the effective interaction between the spin degrees of freedom 
of the {\it opposite} edges finite in the long wavelength limit.
This effect is absent in zero magnetic field and is unique to partially
spin-polarized edges.

We adopt the following model for a narrow Hall bar (or equivalently a
quantum wire 
in strong magnetic fields \cite{YG}).
When the transverse motion is confined
by a parabolic potential the single-electron energy levels are given by 
$E_{n,k}=(n+1/2)\Omega_0+\frac{\hbar^2 k^2}{2m} $,
where the enchanced longitudinal mass is $m=m^* (\Omega_0/\omega_0)^2$. 
The subband energy spacing is  
$\Omega_0=\sqrt{\omega_c^2+\omega_0^2}$, 
where $\omega_c=eB/m^*c$ is the cyclotron energy
and $\omega_0$ is the frequency of the harmonic potential.
In this model, the 
degree of spin-splitting, i.e. the distance 
between the guiding centers of spin-up and -down electrons at the
Fermi wavevectors \cite{DGH}, can be easily tuned by the magnetic field since
the effective longitudinal mass of the electron
depends on the value of the magnetic field \cite{Pinc}.
The electronic wavefunction at
the Fermi wavevectors $k_{F,r,s}$ of the lowest magnetic subband 
is given by 
\begin{equation}
\phi_{r,s}(x,y)=\frac{ e^{i k_{F,r,s} x}}{\pi^{1/4}\,\tilde{\ell}^{1/2}}\,
\exp\Big(-\frac{(y-R^2 k_{F,r,s})^2}{2\tilde{\ell}^2} \Big),
\end{equation}
where  
$\,\tilde{\ell}=(\hbar/m \Omega_0)^{1/2} \,$,
$R^2=\hbar \omega_c /( m \Omega_0^2) \,$, 
$r=R (L)$ for the right (left) branch,  
and $y$ is the transverse coordinate \cite{CP}. 
Because of the large 
longitudinal mass at fields where
$\omega_c> \omega_0$ all the electrons can be accommodated 
in the lowest magnetic subband. 
Each branch $r$ consists of 
spin-up and -down edges ($s=\uparrow,\downarrow$).
The intra branch electron-electron interactions are given by
\ba
& &V^{{\rm intra}}_{q,\parallel}=
-\frac{ 2 e^2}{\epsilon}\,\ln \frac{\gamma q a_{\parallel}}{2},\;\;
V^{{\rm intra}}_{q,\perp}=
-\frac{ 2 e^2}{\epsilon}\,\ln \frac{\gamma q a_{\perp}}{2}
\ea
and the inter branch electron-electron interactions are given by
\ba
& &V^{{\rm inter}}_{q,\parallel,s}=
\frac{ 2 e^2}{\epsilon}\,K_0(q W_{\parallel,s}),\;\;
V^{{\rm inter}}_{q,\perp}=
\frac{ 2 e^2}{\epsilon}\,K_0(q W_{\perp}).
\ea
Here $W_{\parallel,s}=|k_{F,R,s}-k_{F,L,s}|R^2$ and
$W_{\perp}=|k_{F,R,s}-k_{F,L,-s}|R^2$, and 
$K_0(x)$ is 
the modified Bessel function, which behaves like 
$-\ln x$ in the limit of small $x$.
The constants $a_{\parallel}$ and $a_{\perp}$ are the length scales
comparable to the width of the quantum wire, and $v_{F,s}$ are 
the Fermi velocities \cite{com1}. 

It is convenient to use charge and spin 
density operators 
$\, \rho_{r,q}=\frac{1}{\sqrt{2}}\,\Big(\rho_{r,q,\uparrow}+
\rho_{r,q,\downarrow} \Big)$ and
$\sigma_{r,q}=\frac{1}{\sqrt{2}}\,\Big(\rho_{r,q,\uparrow}-\rho_{r,q,\downarrow} \Big)
\,$.
The bosonized Hamiltonian can be written as a sum of the charge, spin, 
and mixed terms
$ H=H_\rho+H_\sigma+\delta H $,
where
\begin{eqnarray}
& &H_\rho=\frac{\pi v_F}{L}\,\sum_{r,q \neq 0}\,
: \rho_{r,q}\,\rho_{r,-q}:+ \nonumber \\
&&\frac{1}{L}\,\sum_{r,q}\,V^{{\rm intra}}_{\rho,q}\,:\rho_{r,q}\,\rho_{r,
-q}: 
+ \frac{2}{L}\,\sum_q  V^{{\rm inter}}_{\rho,q}\, \rho_{R,q}\,\rho_{L,-q},
\end{eqnarray}
\begin{eqnarray}
& &H_\sigma=\frac{\pi v_F}{L}\,\sum_{r,q \neq 0}\,
:\sigma_{r,q}\,\sigma_{r,-q}:+ \nonumber \\
&&\frac{1}{L}\,\sum_{r,q}\,V^{{\rm intra}}_{\sigma,q}\,
:\sigma_{r,q}\,\sigma_{r,-q}: 
+\frac{2}{L}\,\sum_qV^{{\rm inter}}_{\sigma,q}\,\sigma_{R,q}\,\sigma_{L,-q},
\end{eqnarray}
and
\begin{eqnarray}
& &\delta H=2 \delta v_F\,\frac{\pi}{L}\,\sum_{r,q}\,\sigma_{r,q}\,\rho_{r,-q}
\nonumber \\
&+&\frac{1}{L}\,\sum_q\,\delta V^{{\rm inter}}_q\,\Big[\sigma_{R,q} \rho_{L,-q}
+\sigma_{L,q}\, \rho_{R,-q} \Big],
\end{eqnarray}
with
\begin{eqnarray}
& &V^{{\rm intra}}_{\rho (\sigma),q}=
\frac{1}{2}\Big(V^{{\rm intra}}_{q,\parallel}\pm
V^{{\rm intra}}_{q,\perp}\Big),\;\;\nonumber \\
& &V^{{\rm inter}}_{q,\parallel}
=\frac{1}{2}\,\Big(V^{{\rm inter}}_{q,\parallel,\uparrow}
+V^{{\rm inter}}_{q,\parallel,\downarrow}\Big),\;\; \nonumber \\
& &\delta V^{{\rm inter}}_q=
\frac{1}{2}\,\Big(V^{{\rm inter}}_{q,\parallel,\uparrow}
-V^{{\rm inter}}_{q,\parallel,\downarrow}\Big), \nonumber \\
& &V^{{\rm inter}}_{\rho (\sigma),q}=
\frac{1}{2}\Big(V^{{\rm inter}}_{q,\parallel}\pm
V^{{\rm inter}}_{q,\perp}\Big),\;\; \nonumber \\
& &v_F=\frac{1}{2}\,\Big(v_{F\uparrow}+v_{F \downarrow} \Big),\;\;
\delta v_F=\frac{1}{2}\,\Big(v_{F\uparrow}-v_{F \downarrow} \Big) .
\end{eqnarray}
Note that in the effective Hamiltonian, 
the charge and spin degrees of freedom are 
separated  
except in $\delta H$ \cite{LY}.

In the computation of correlation functions the {\em phase fields} 
($\,\theta_{\rho,\sigma},\;\;\phi_{\rho,\sigma} \,$ ) formulation 
is more convenient. In momentum space, they are defined as
\begin{eqnarray}
& &-\frac{i q}{\pi}\, \phi_{\rho}=\rho_R+\rho_L,\;\;
-\frac{i q}{\pi}\, \phi_{\sigma}=\sigma_R+\sigma_L, \nonumber \\
& &\frac{i q}{\pi} \,\theta_{\rho}=\rho_R-\rho_L,\;\;
\frac{i q}{\pi} \,\theta_{\sigma}=\sigma_R-\sigma_L.
\end{eqnarray}
We can derive the effective action corresponding to 
the Hamiltonian by
the standard procedure. The action in imaginary time and 
at zero temperature is
\begin{eqnarray}
\label{action}
S&=&\frac{1}{2\pi}\,\int \,\frac{d\omega dk}{(2\pi)^2}\,
\Big[(\omega^2\,\frac{v_{\sigma -}}{v_{\rho -} v_{\sigma -}-g_-^2}
+v_{\rho +} k^2 ) \,\phi_{\rho} \phi_{\rho} \nonumber \\
&+&( \omega^2\,\frac{v_{\rho -}}{v_{\rho -} v_{\sigma -}-g_-^2}+
v_{\sigma +} k^2) \,\phi_\sigma \phi_\sigma  \nonumber \\
&+& 2 (
-\omega^2\,\frac{g_-}{v_{\rho -} v_{\sigma -}-g_-^2}+g_+ k^2)\, \phi_\sigma 
\phi_\rho  \Big],
\end{eqnarray}
where 
\begin{eqnarray}
v_{\rho \pm}&=&v_F+\frac{V^{{\rm intra}}_{\rho,q}}{\pi}\pm
\frac{V^{{\rm inter}}_{\rho,q}}{\pi},\quad \nonumber \\
v_{\sigma \pm}&=&v_F+\frac{V^{{\rm intra}}_{\sigma,q}}{\pi}
\pm\frac{V^{{\rm inter}}_{\sigma,q}}{\pi},\quad \nonumber \\
g_{\pm}&=& \delta v_F \pm \frac{\delta V^{{\rm inter}}_q}{\pi}.
\end{eqnarray}
For later convenience, we define $\,W_\parallel=\Big(W_{\parallel \uparrow} \,
W_{\parallel \downarrow} \Big)^{1/2} \,$ and 
$ \, v_0=\frac{ 2 e^2 }{\hbar \pi \epsilon } \,$. 
The  Eq.(\ref{action}) can be also expressed in terms of conjugate phase fields 
%$\, \theta_{\rho,\sigma}(x) \,$,
by simply replacing $v_{\rho \pm} \to v_{\rho \mp},\,
v_{\sigma \pm} \to v_{\sigma \mp}$, and  $g_{\pm} \to g_{\mp} \,$.
We will use the action (\ref{action}) and its conjugate action for the 
computation of  correlation 
functions. The propagators of phase fields can be obtained by inverting 
kernel matrices of  (\ref{action})  and its conjugate action.
Finally, we need the explicit expression 
of electron operators in terms of phase fields \cite{zulicke96}
\begin{equation}
\psi_{r,s}(x,y)=\phi_{r,s}(x,y)\,e^{-\frac{i}{\sqrt{2}}\,
\Big(r(\phi_\rho+s \phi_\sigma)+(\theta_\rho+s \theta_\sigma)\Big)}. 
\end{equation}
It is slightly different from the bosonization formula of a 
truly 1D system because
the left and right edges are spatially separated by the width of the
quantum wire. 

Let us consider the correlation function of the transverse spin operator.
The transverse spin operator is 
\begin{eqnarray}
\label{spinoperator}
& &\hat{S}^{+}(x)=\int dy \,\sum_{r,r^{\prime}=R,L}
 \,\psi^{\dag}_{r \uparrow}(x,y)\,
 \psi_{r^{\prime}\downarrow}(x,y)   \nonumber \\
&=&
C_0(x)\,e^{-i \sqrt{2} \theta_\sigma (x)}\,\cos \sqrt{2} \phi_\sigma (x)
\nonumber \\
&+&C_{2 k_F}(x) \,
e^{-i \sqrt{2}\,\theta_\sigma(x)}\,\cos \sqrt{2} \phi_\rho(x),
\end{eqnarray}
where $\,C_{0 (2k_F)}(x)=e^{-i(k_{F\uparrow}\mp k_{F \downarrow}) x}\,
e^{-\frac{(k_{F\uparrow}\mp k_{F \downarrow})^2 R^4}{4 \tilde{\ell}^2}}$.
The first term  of Eq.(\ref{spinoperator}) 
is the intra-branch contribution, and the second term is the inter-branch
contribution ($ 2 k_F $ component).
Note that the intra-branch spin operator is composed entirely of spin bosons 
($\phi_\sigma (x)$ and $\theta_\sigma (x)$) while the 
inter-branch spin operator is composed of
both the charge $\phi_\rho (x)$ and
spin $\theta_\sigma (x)$ degrees of freedom.
When the off-diagonal elements
of the action are negligible
the correlation function of the intra-branch spin operator
will only reflect the spin degree of freedom.  

The inter- and intra-branch terms of the imaginary part of the 
transverse correlation function are  
given in
Eqs. (\ref{answer}) and (4)\cite{com2}.
The the
anomalous exponent $\alpha_{2 k_F}$ is given by \cite{com3}
\begin{equation}
\label{2kf}
\alpha_{2 k_F}=\left[ \frac{ \frac{v_F}{ v_0}+\frac{1}{2}\,
\ln \frac{a_\perp W_\perp}{a_\parallel W_\parallel}
}{\frac{v_F}{ v_0}+\frac{1}{2}\,
\ln \frac{a_\perp W_\parallel}{a_\parallel W_\perp} }\right]^{1/2}
\,-1.
\end{equation}
The other anomalous exponent $\alpha_0$ is
\begin{equation}
\alpha_0=\left[ \frac{ \frac{v_F}{ v_0}+\frac{1}{2}\,
\ln \frac{a_\perp W_\perp}{a_\parallel W_\parallel}
}{\frac{v_F}{ v_0}+\frac{1}{2}\,
\ln \frac{a_\perp W_\parallel}{a_\parallel W_\perp} }\right]^{1/2}
\, +[ W_\parallel \leftrightarrow  W_\perp ]^{1/2}-2.
\end{equation}
%The other constants appearing in Eq.(\ref{answer}) are
%\begin{eqnarray}
%\,v_\rho&=&\Big[ v_0 (2 v_F+v_0\,\ln
%\frac{W_\parallel W_\perp}{a_\parallel a_\perp})
% \Big]^{1/2} \,\nonumber\\
%\,d&=&\frac{\gamma e^{-v_F/(2v_0)}}{2}\,
%\Big( W_{\parallel} W_\perp a_\parallel a_\perp
% \Big)^{1/4} \, \nonumber\\
%\beta_{2k_F}&\sim& \sqrt{2} \Big[\frac{v_F}{v_0}
%+\frac{1}{2}\,\ln \frac{W_\parallel W_\perp}{
% a_\parallel a_\perp} \Big]+\frac{ g_\pm^2}{v^2_{\rho}}
%\end{eqnarray}
The crossover frequency $\Omega_{{\rm cr}}$
is roughly 
$ v_{\rho}/\sqrt{W_\parallel W_\perp} $ and the anomalous exponents are 
always non-negative since
$\frac{W_\perp}{W_\parallel}>1$.
The amplitude of the $2 k_F$ component of correlation 
function  $|C_{2 k_F}(x)|^2$ is explicitly given by
$e^{-2 R^4 k_F^2 /\tilde{\ell}^2}=\exp\Big[-4\,\frac{E_F}{\omega_0}\,
\frac{\omega_0 \omega_c^2}{\Omega_0^3}\Big]$.
For ordinary QH bars this amplitude is negligibly small \cite{zulicke96},
but for a narrow QH bar or a quantum wire in strong magnetic fields
it is of order one for reasonable values of $E_F, \omega_0$, and $B$. 
For reasons given in the second paragraph $\alpha_{2k_F}$
is experimentally more relevant than $\alpha_0$.  
From Eq. (\ref{2kf}) we see that 
when $V_{\sigma,q}^{{\rm inter}}=0$ ( $ W_\parallel=W_\perp $)
the anomalous exponent $\alpha_{2 k_F}$ vanishes irrespective
 of $a_\parallel$ and $a_\perp$. 
This means only the $V_{\sigma,q}^{{ \rm inter}}$ interaction 
in the spin part of the Hamiltonian
$H_\sigma$
can give rise to an anomalous exponent.  In the long wavelength limit
this interaction takes a {\it finite} value
since it is given by the {\it difference} between two modified Bessel
functions.
In contrast the effective interactions 
$V_{\rho,q}^{{\rm intra}}$ and
$V_{\rho,q}^{{\rm inter}}$ of $H_{\rho}$ diverge in the same limit.  
The numerical value of $\alpha_{2k_F}$ is dependent on the detailed 
shape of the confining potential, and therefore, is not universal.

The correlation function of the longitudinal spin operator is very similar to 
that of transverse one. The same crossover exists, and 
the inter-branch contribution dominates over the intra-branch
contribution in the Coulomb regime. The anomalous exponent
$\alpha_{2 k_F}$ is obtained
by replacing $W_\perp$ with $W_\parallel$ in Eq. (17)\cite{com4}. 
Because $ W_\perp > W_\parallel$ the exponent
$ \alpha_{2 k_F} $ is now negative, 
which signals the SDW -like ground state at zero temperature.  Recall 
that in our system the spin SU(2) symmetry is 
broken by magnetic field, 
and it is natural that the transverse and longitudinal 
spin correlation functions have  different  anomalous exponents.
We have also calculated the cross-over behavior of charge correlation
functions, and find that 
it is almost identical with those of 
quantum wires at zero magnetic field and spin-polarized edges 
at  filling factor 1 \cite{FB,schulz93}.
In quantum wires one can observe upon bosonization that
the long-range
Coulomb interaction couples only to the charge 
degree of freedom \cite{schulz93}.
In this case an anomalous exponent is absent in the spin sector.

We now discuss the experimental relevance of
the frequency dependence of the transverse spin correlation function.  
We believe that NMR measurements of a quantum wire in strong magnetic 
fields should demonstrate 
the presence of the anomalous exponent since
the transverse spin correlation function
is directly related to the nuclear spin relaxation rate.
(NMR measurements of quantum Hall edges have been carried out 
recently \cite{wald94}).
The effect of the spin-splitting on 
the anomalous exponent would be most strong 
when the separation between spin-up and -down Fermi edges is large.
In such a system the widths
$W_{\parallel}$ and $W_\perp$ would be rather different and should yield
a significant value of the anomalous exponent.  To get an estimate of 
$\alpha_{2 k_F}$, we expand Eq.(\ref{2kf}), and find 
$\alpha_{ 2 k_F} \approx \frac{ 4 v_0}{v_F}\,\ln
 \frac{W_\perp}{W_\parallel} $.  Since
$v_F$ is of order $v_0$ in the absence of a magnetic field 
it can be made significantly smaller than $v_0$ by applying a strong magnetic 
field \cite{com1}.
>From this we estimate that $\alpha_{2 k_F} \sim 0.1$ is reasonable.

In conclusion, we have shown that
the charge and spin correlation functions of 
spin-polarized edge states 
behave qualitatively differently.  This effect is unique to 1D systems
in strong magnetic fields and
is based on the novel property of shifting of the guiding center
of the Landau level wavefunction with the change in the 
single-particle quantum number.  In the long wavelength limit
the effective interaction for the spin degree
of freedom takes a finite value while
the effective interaction for the charge degree
of freedom is infinitely strong.  As a consequence, an anomalous exponent
appears in the spin sector.  The presence of the anomalous exponent may
be tested experimentally in a narrow QH bar or in a 
quantum wire in strong magnetic fields.

H. C. Lee is grateful to J. H . Han for useful comments.  S.R.E.Y has been
supported by the KOSEF under grant 961-0207-040-2 and the Ministry of
Education under grant BSRI-96-2444.

\end{multicols}

\begin{references}
\bibitem{com1} The Fermi energy of a typical quantum wire in a strong magnetic
field is of order 1 meV, which gives $v_F=7\times10^6 {\rm cm/s}$.  
This value should be
compared with $v_0=1.3\times 10^7 {\rm cm/s}$.
\bibitem{com2} 
%The constant  
%$\beta_0 \sim \frac{ g_{\pm}^2}{v^2_{\rho,\sigma}}$ ,
%and originates entirely
%from the mixed term $\delta H$ in the Hamiltonian. 
Compared with the power law of the inter-branch part
the intra-part has the additional power of 1 in $\Omega$. It stems from
the difference of $\,e^{-2 \langle 
\phi_\sigma \phi_\sigma \rangle} \,$ and $
\,e^{-2 \langle \phi_\rho \phi_\rho \rangle} \,$ in the Coulomb
regime. The former 
yields $1/{|\tau|}^{1+\tilde{\alpha}}$, while the latter
gives $e^{-\sqrt{\ln |\tau |}}$.
\bibitem{com3} When the mixed term $\delta H$ is significant
the square root must be multiplied by $1+\delta g+ O(\delta g^2)$, where 
$\delta g = \frac{1}{2}\,\frac{g_-^2}{
(v_F+\frac{1}{2}\,v_0\,\ln \frac{W_\parallel W_\perp}{a_\parallel a_\perp})
(v_F+\frac{1}{2}\,v_0\,\ln \frac{W_\parallel a_\perp}{a_\parallel W_\perp})}$.
In quantum wires $\delta g$ is always less than unity,
which justifies the expansion in $\delta g$.
\bibitem{com4}When the mixed term $\delta H$ is significant
the square root must be mutiplied by $1-\delta g$.
\end{references}
\end{document}